\documentclass{article}
\usepackage{spconf,amsmath,graphicx}
\usepackage{multirow}
\usepackage{subcaption}
\usepackage[boxsize=2.6em]{ytableau}
\usepackage{multicol}
\ninept

\title{Multi-Band Multi-Resolution Fully Convolutional Neural Networks for Singing voice Separation}
\name{Emad M. Grais$^{1}$\sthanks{Emad is on unpaid leave from Helwan University, Egypt.}, Fei Zhao$\,^{1}$, Mark~D.~Plumbley$^{2}$}
\address{
$^1$Centre for Speech and Language Therapy and Hearing Science, Cardiff Metropolitan University, UK \\
$^2$Centre for Vision, Speech and Signal Processing, University of Surrey, UK
}
\begin{document}
%
\maketitle
\begin{abstract}
Deep neural networks with convolutional layers usually process the entire spectrogram of an audio signal with the same time-frequency resolutions, number of filters, and dimensionality reduction scale. According to the constant-Q transform, good features can be extracted from audio signals if the low frequency bands are processed with high frequency resolution filters and the high frequency bands with high time resolution filters. In the spectrogram of a mixture of singing voices and music signals, there is usually more information about the voice in the low frequency bands than the high frequency bands. These raise the need for processing each part of the spectrogram differently. In this paper, we propose a multi-band multi-resolution fully convolutional neural network (MBR-FCN) for singing voice separation. The MBR-FCN processes the frequency bands that have more information about the target signals with more filters and smaller dimentionality reduction scale than the bands with less information. Furthermore, the MBR-FCN processes the low frequency bands with high frequency resolution filters and the high frequency bands with high time resolution filters. Our experimental results show that the proposed MBR-FCN with very few parameters achieves better singing voice separation performance than other deep neural networks.  
\end{abstract}
\section{Introduction}
\label{sec:intro}
Separating the singing voice from a single mixture of music and vocal signals has many applications, such as soloing, karaoke, and remixing for hearing assistive devices \cite{Simpson:15:dkevmmcdnn,Ponsa:16:rmssaimeciu,Gajeckia:18:dlmrciu}. Many deep learning models, such as deep feed forward neural networks (DNN), convolutional neural networks, recurrent neural networks, and fully convolutional neural networks (FCN) have been proposed for single channel singing voice separation (SCSVS) \cite{Grais:14:dnnscss,roma:18:isnscsmacl,Huang:14:svsmrdrnn,grais:18:mrfcnnmass}. In all these models, the neural network processes the entire spectrogram of an audio signals with the same degree of importance. In the spectrogram of a mixture of many audio sources, some sources in the mixture might have more information at certain frequency bands (important bands) than other bands. This means we need to extract more details about the target sources in the important bands than the other bands. Unlike images where different patterns can appear anywhere in the image, the spectrograms of audio signals have different patterns at different frequency bands \cite{Huang:01:slpgtasd,Gold:11:saspppsm}. In addition, depending on the application (classification or regression), it is not always suitable to apply dimensionality reduction with the same scale over the extracted features from the entire spectrogram as the case with images, since in some bands of the spectrogram we might need to keep all the extracted features, while in other bands we might reduce the dimensionality of the extracted features. Following the constant-Q transform \cite{Brown:91:ccqst}, good features could be extracted from audio signals if the low frequency bands are processed by high frequency resolution filters and the high frequency bands are processed by high time resolution filters. These suggest that different bands of frequencies should be processed differently. In \cite{Takahashi:17:msmbass}, a combination of two neural networks was introduced, where the first neural network processes the entire spectrogram and the second neural network processes the input spectrogram as two separate frequency bands.

In this paper, we propose a multi-band multi-resolution fully convolutional neural network (MBR-FCN) for single channel singing voice separation (SCSVS). The spectrogram of the input mixture is divided into overlapped frequency bands following the same concept of computing the Mel-frequency cepstral coefficients (MFCCs) features, which are considered one of the best audio features and have been used in many audio applications \cite{Davis:80:cprmwrcss}. The proposed MBR-FCN processes each band of frequencies in the spectrogram with different degree of importance depends on how much information we expect in each band. The bands with more information are processed with more filters and small dimensionality reduction scale is applied on the extracted features, while the bands with less information are processed with few filters and high dimensionality reduction scale is applied on the extracted features. Following the constant-Q transform \cite{Brown:91:ccqst}, the proposed MBR-FCN processes the low frequency bands with high frequency resolution filters and the high frequency bands with high time resolution filters. The proposed MBR-FCN has very few parameters.  

This paper is organized as follows. In Section \ref{sec:mbmrfcnn}, the details of the proposed MBR-FCN are given. In section \ref{sec:experiment}, the experiments and results are shown. In the remaining section, the conclusion is presented.
%
\section{Multi-Band Multi-Resolution Fully Convolutional Neural Networks}
\label{sec:mbmrfcnn}
\begin{figure}[t!]
\begin{center}
 \includegraphics[width=1\linewidth,height=6cm]{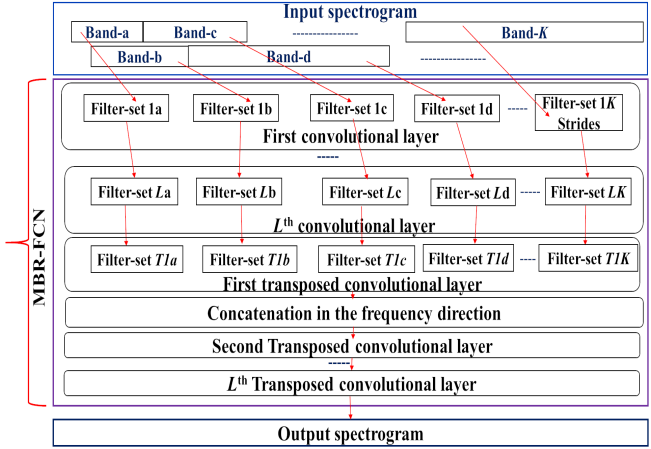}
\caption{\label{fig:layer_MB_FCN_basic}{The proposed multi-band multi-resolution fully convolutional neural network (MBR-FCN). It has $L$ convolutional layers followed by $L$ transposed convolutional layers. The input spectrogram is divided into $K$ overlapped frequency bands. Strides are not used in the bands that could have information about the target signals.}}
\end{center}
\end{figure}

Fig. \ref{fig:layer_MB_FCN_basic} shows the proposed multi-band multi-resolution fully convolutional neural network (MBR-FCN). The MBR-FCN has $L$ convolutional layers followed by $L$ convolutional transposed layers, where all layers are composed of 2D filters. As shown in Fig. \ref{fig:layer_MB_FCN_basic}, the spectrogram of the input mixture is divided into $K$ overlapped frequency bands similar to the way the MFCCs are computed \cite{Davis:80:cprmwrcss}. The bands are spaced equally in the mel-frequency scale \cite{Huang:01:slpgtasd}. The mel-frequency scale is computed from the linear scale as follow:
\begin{equation}
\label{mel_scale}
M(f) = 1125\ln{ (1+f/700) },
\end{equation}
where $f$ is the frequency value in the linear frequency scale of the spaectrogram and $M(f)$ is its corresponding value in the mel-scale. 

\begin{figure}[t!]
\begin{minipage}[b]{1.0\linewidth}
  \centering
  \centerline{\includegraphics[width=.95\linewidth,height=5.5cm]{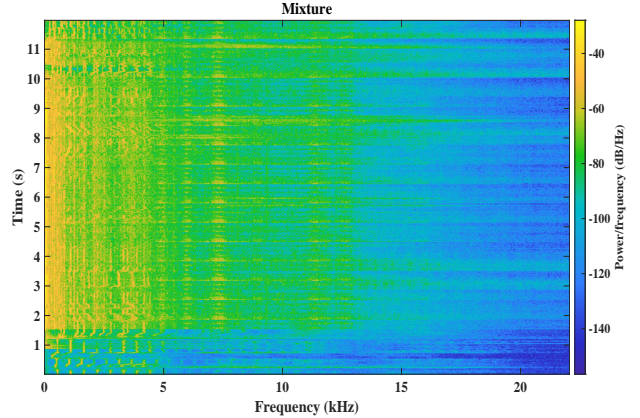}}
  \centerline{(a) A mixture of singing voices and music signals}\medskip
\end{minipage}
\begin{minipage}[b]{1.0\linewidth}
  \centering
  \centerline{\includegraphics[width=.95\linewidth,height=5.5cm]{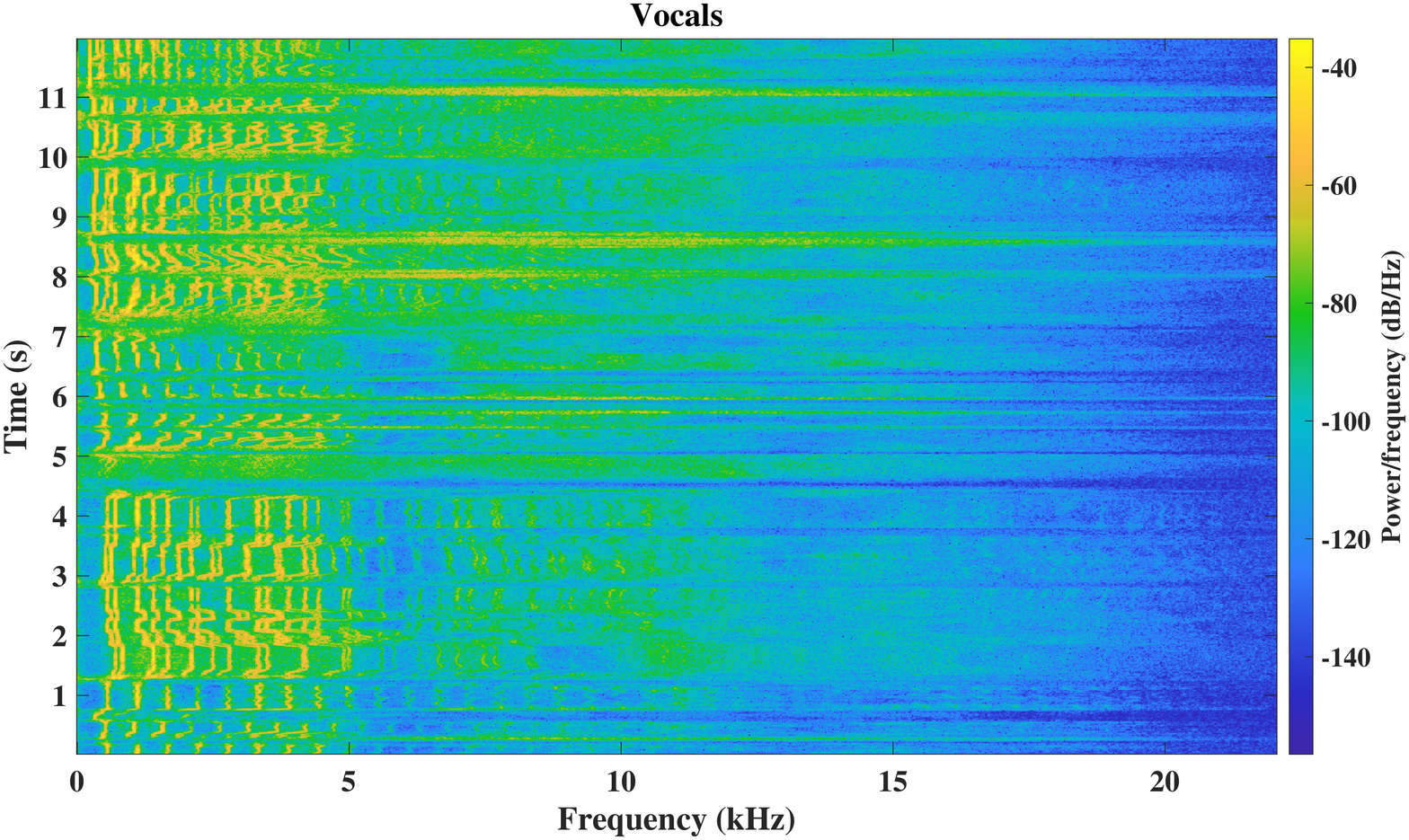}}
  \centerline{(b) Singing voices}\medskip
\end{minipage}

\caption{Spectrogram examples for (a) a mixture of singing voices and music, (b) the corresponding singing voices.}
\label{fig:spectrogram}
\end{figure}

 Fig. \ref{fig:spectrogram}-(a) shows an example of the spectrogram of a mixture of singing voices and music signals, while Fig. \ref{fig:spectrogram}-(b) shows the spectrogram of the corresponding target singing voices in the mixture. The data and parameters for computing these spectrograms will be discussed in Section \ref{sec:experiment}. As shown in the spectrograms in Fig.\ref{fig:spectrogram}, most of the information related to the singing voices is in the low frequency bands. Furthermore, the information is more dense at low frequency bands than at high frequency bands. These suggest that more details about the target signals could be extracted by processing the low frequency bands with more filters than the high frequency bands. Because of the dense information at low frequencies, the sets of the 2D filters in the MBR-FCN that process the low frequency bands of the spectrogram should capture frequency information in high frequency resolution. Since the low frequency bands has more information than the high frequency bands, the scale of the dimensionality reduction that is applied at the extracted features from the low frequency bands should be smaller than the scale of dimensionality reduction that is applied at the extracted features from the high frequency bands. The proposed MBR-FCN reduces the dimensionality (using strides \cite{Dumoulin:16:agcadl}) in the frequency direction for the extracted features from the bands where we do not expect to find important information related to the target signals, thus MBR-FCN uses strides at high frequency bands as shown in Fig. \ref{fig:layer_MB_FCN_basic}.

We can also see from the spectrograms in Fig. \ref{fig:spectrogram} that the patterns at low frequency bands are longer in the time direction than the patterns at high frequency bands, this suggests that the 2D filters at low frequency bands should be longer (to correlate with the size of the patterns) in the time direction at low frequency bands than at high frequency bands. The short 2D filters in the time direction at high frequency bands also help in extracting features with high time resolution at high frequency bands. Extracting features with high frequency resolution at low frequency bands and high time resolution at high frequency bands in this work follows the constant-Q transform \cite{Brown:91:ccqst}.   
%

Each filter set in the MBR-FCN is followed by batch normalization and a nonlinear activation function. To consider the mutual information among all the bands, the extracted features from all bands (that are extracted based on different degree of importance) are then concatenated in the frequency axis at the transposed convolutional layers. The MBR-FCN is trained to predict the entire magnitude spectrogram of the target singing voices in its output. The learning of the MBR-FCN parameters is done by minimizing the mean-square-errors between the estimated spectrogram and the reference spectrogram for the singing voices. 
%
%
%
%
%

\section{Experiments}
\label{sec:experiment}
We used our proposed MBR-FCN to separate the singing voice/vocal sources from a group of songs from the SiSEC-2015-MUS-task dataset \cite{ono:15:tsisec}. The dataset has 100 stereo songs with different genres and instrumentations. The stereo songs were converted to mono by averaging the two channels. Each song is a mixture of vocals, bass, drums, and a group of other musical instruments. 

The first 50 songs in the dataset were used as training and validation datasets, and the last 50 songs were used for testing. The data were sampled at $44.1$\,kHz. The magnitude spectrograms for the data were calculated using the STFT with Hanning window size 2048 points and hop size of 512 points. The FFT was computed with 2048 points and the first 1025 were used as features since they include the conjugate of the remaining points. Fig. \ref{fig:spectrogram} shows the spectrograms of a segment of one of the songs in the test set.

The spectrograms of the input signals are divided into five overlapped frequency bands ($a,b,c,d,e$ in Fig. \ref{fig:layer_MB_FCN_basic}). The frequency range in each band as FFT indices and Hz is shown in Table \ref{table:bins}. The number of bins in some bands is slightly modified from the way MFCCs are usually computed \cite{Davis:80:cprmwrcss} to make the dimension of the features after the concatenation layer to be 1025. Note that, strides with value three is used in band-e after the first convolution layer.
\begin{table}[t!]
\centering
\caption{The range of frequencies as FFT indices and Hz in each frequency band in Fig. \ref{fig:layer_MB_FCN_basic}.}
\label{table:parmtrs}
\begin{tabular}{|c|c|c|c|c|}
\hline
\multirow{2}{*}{Bands} & \multicolumn{2}{c|}{frequencies as FFT indices} & \multicolumn{2}{c|}{frequencies in Hz}  \\
                       & From                  & To                    &          From         & To \\
\hline
a & 0   & 73    &  0   & 1572 \\
b & 26  & 156   & 560  & 3360  \\
c & 73  & 305   & 1572 & 6568 \\
d & 221 & 571   & 4759 & 12295 \\
e & 305 & 1025  & 6568 & 22050  \\
\hline
\end{tabular}
\label{table:bins}
\end{table}

For the MBR-FCN, it has two convolutional layers and two transposed convolutional layers, thus $L=2$ in Fig. \ref{fig:layer_MB_FCN_basic}. The number and size of the filters that process each band in each layer are shown in Table \ref{table:models}. We can see from Table \ref{table:models} that within the same convolutional layer the size of the filters in the time direction is decreasing starting from the low frequency bands to the high frequency bands, while the size of the filters in the frequency direction is increasing starting from the low frequency bands to the high frequency bands. These filters with small size in the time direction at high frequency band extract features in high resolution in time, and the filters with small sizes in the frequency direction extract features in high frequency resolution in the low frequency bands as disscused in Section \ref{sec:mbmrfcnn}.

As shown in Fig. \ref {fig:spectrogram}-(b), most of the information of the singing voice is usually under $5$\,kHz and this is covered in the first three bands in the spectrograms. Thus, we used few filters in the fourth and fifth bands in the convolutional layers in the MBR-FCN as shown in Table \ref{table:models}. In the first convolutional layer, we used strides in the frequency direction in the last band since we do not expect important information about the vocal in the last band, but we still interested to extract some important information from the background signals that might help with the separation process. The value of the strides used in the last band is three in the frequency direction only.
%
%
%
%
%
\begin{figure}[h!]
\begin{center}
 \includegraphics[width=1\linewidth,height=4.5cm]{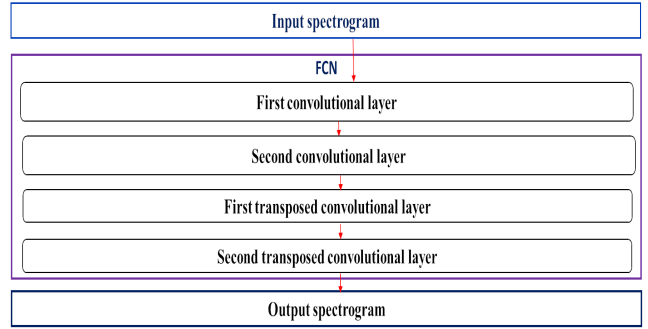}
\caption{\label{fig:fcn}{\scriptsize{The fully convolutional neural network (FCN).}}}
\end{center}
\end{figure}
\begin{table}[h!]
\begin{center}
\scalebox{0.9}
{
\begin{tabular}{|p{1.3cm}|p{3cm}|p{.75cm} p{2.9cm}|}
 \hline
 \multicolumn{4}{|c|}{FCN and MBR-FCN model summary} \\
 \hline
 \multicolumn{4}{|c|}{The input/output data with size 29 frames and 1025 frequency bins} \\
 \hline
Layer No. & FCN & \multicolumn{2}{|c|}{MBR-FCN} \\
  \hline
\multirow{3}{*}{1} & \multirow{5}{*}{Conv2D[25,(11,42)]}    & set 1a & Conv2D[7,(15,11)]\\
                   &                                        & set 1b & Conv2D[7,(13,18)]\\ 
                   &                                        & set 1c & Conv2D[5,(11,33)]\\
                   &                                        & set 1d & Conv2D[3,(9,51)]\\
                   &                                        & set 1e & Conv2D[3,(7,101)]\\
\hline 
\multirow{3}{*}{2} & \multirow{5}{*}{Conv2D[55,(11,22)]}    & set 2a & Conv2D[15,(15,7)]\\
                   &                                        & set 2b & Conv2D[15,(13,11)]\\ 
                   &                                        & set 2c & Conv2D[13,(11,19)]\\
                   &                                        & set 2d & Conv2D[6,(9,25)]\\
                   &                                        & set 2e & Conv2D[6,(7,51)]\\
\hline
\multirow{3}{*}{3} & \multirow{5}{*}{Conv2DTr[25,(15,131)]} &  set 3a & Conv2DTr[5,(15,131)]  \\
                   &                                        &  set 3b & Conv2DTr[5,(15,131)] \\ 
                   &                                        &  set 3c & Conv2DTr[5,(15,131)] \\
                   &                                        &  set 3d & Conv2DTr[5,(15,131)] \\
                   &                                        &  set 3e & Conv2DTr[5,(15,131)] \\
                   
\hline 
4                  &   Conv2DTr[1,(29,1025)]                &  \multicolumn{2}{|c|}{Conv2DTr[1,(29,1025)]} \\
  \hline
\end{tabular}
}
\caption{The filter specifications and the number of filters in each layer of the FCN and MBR-FCN. For example ``Conv2D[25,(11,42)]'' denotes a 2D convolutional layer with 25 filters and the size of each filter is 11$\times$42 where 11 is the size of the filter in the time-frame direction and 42 in the frequency direction of the spectrogram. The term ``Conv2DTr'' denotes a transposed convolution layer.}
\label{table:models} 
\end{center}
\end{table}
The performance of the proposed MBR-FCN was compared with three different deep neural networks: the deep fully connected neural network (DNN), the fully convolutional neural network (FCN), and the deep U-Net convolutional networks (U-Net) \cite{Jansson:17:svsduncn}. The DNN has four hidden layers, and each hidden layer has 1025 nodes. The details of the FCN is shown in Fig. \ref{fig:fcn} and Table \ref{table:models}. The FCN has the same number of layers as MBR-FCN. The number of filters in each layer in FCN equals to the total number of filters in its corresponding layer in MBR-FCN. The sizes of the filters in each layer in FCN are chosen to be almost equal to the average of the sizes of the filters in all bands in the corresponding layer in MBR-FCN. The U-Net has almost the same structure as the FCN except that the output of the first convolutional layer is concatenated in the channel direction with the output of the first convolutional transposed layer and fed to the second convolutional transposed layer. The rectified linear unit (ReLU) is used as the activation function for all the neural networks in this work. Table \ref{table:parameters} shows the number of parameters in each model. The proposed MBR-FCN model has the fewest parameters compared with the other models. Each input and output segment for the FCN, U-Net, and MBR-FCN is composed of 29 neighbour frames from the input and output spectrograms respectively, which means the size of each input and output segment is 29-frames $\times$ 1025 frequency bins. Each input and output of the DNN is a single frame from the spectrograms of the input mixture and output target source respectively.    %
\begin{table}[h!]
\centering
\caption{The number of parameters (NoP) in each model.}
\label{table:parmtrs}
\begin{tabular}{|c|c|c|c|c|}
\hline
Models   &DNN & FCN & U-Net & MBR-FCN  \\
\hline
NoP      & 4,206,600 & 3,789,506 & 4,532,631  &  747,733 \\
\hline
\end{tabular}
\label{table:parameters}
\end{table}

The parameters for all the models were initialized randomly. They were trained using backpropagation with gradient descent optimization using Adam \cite{adam:14:amso} with parameters $\beta_1=0.9$, $\beta_2=0.999$, $\epsilon=1e-08$, a batch size 100, and a learning rate of $0.0001$, which was reduced by a factor of 10 when the values of the cost function ceased to decrease on the validation set for 3 consecutive epochs. We implemented our proposed algorithm using Keras with Tensorflow backend \cite{site:chollet2015keras3}. 
\begin{figure}[t!]
\begin{minipage}[b]{1.0\linewidth}
  \centering
  \centerline{\includegraphics[width=.9\linewidth, height=5.6cm]{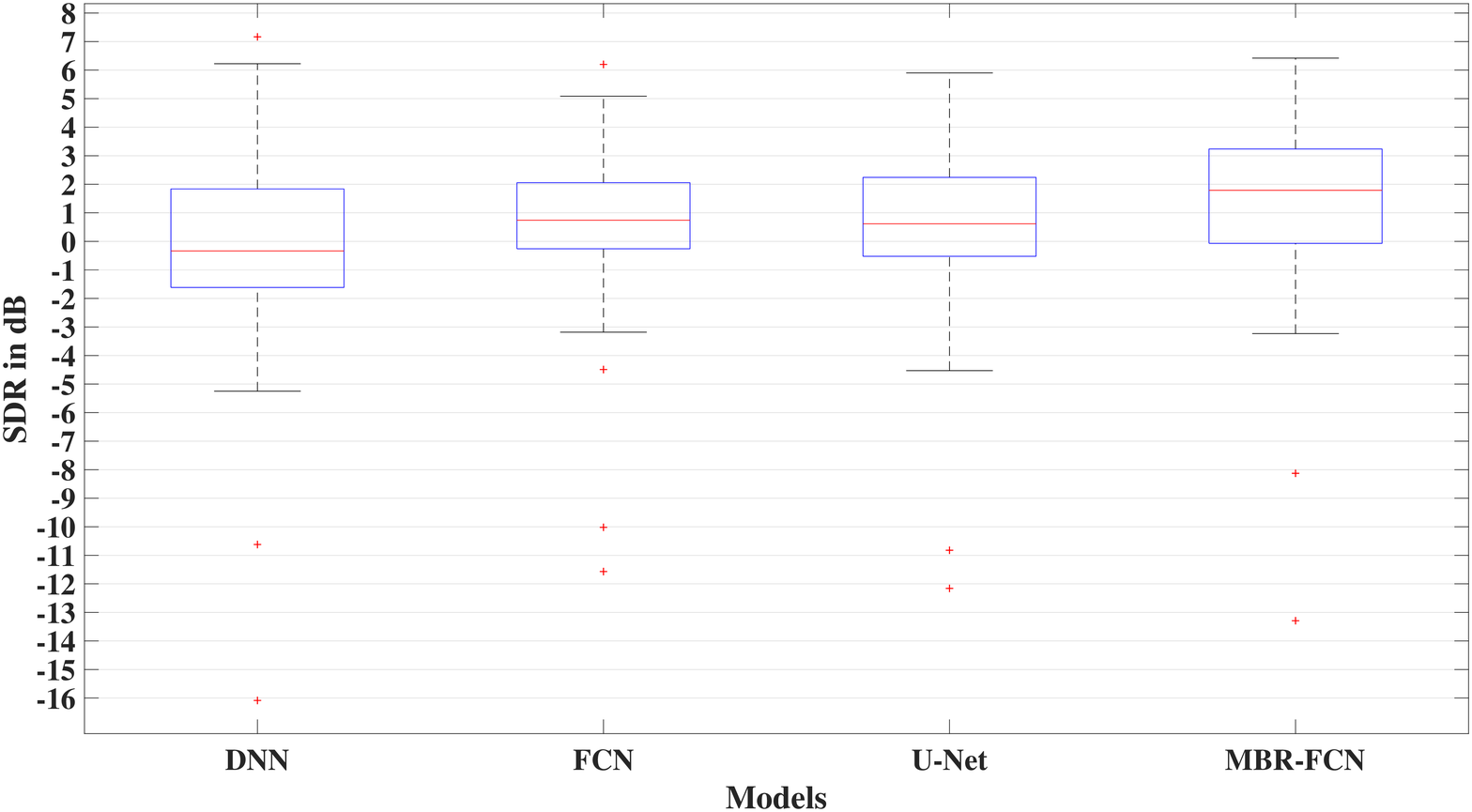}}
  \centerline{(a) SDR}\bigskip
\end{minipage}
\begin{minipage}[b]{1.0\linewidth}
  \centering
  \centerline{\includegraphics[width=.9\linewidth, height=5.6cm]{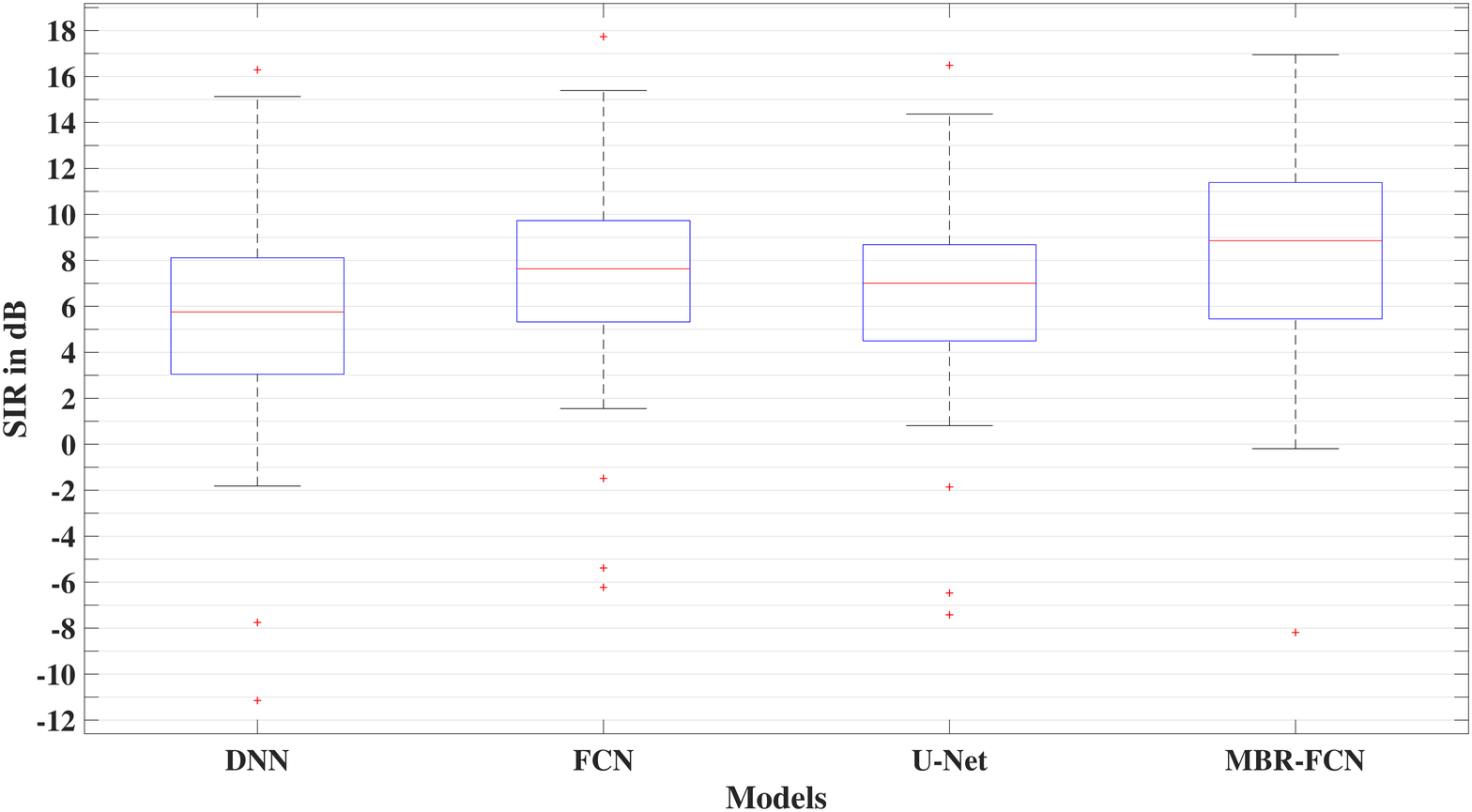}}
  \centerline{(b) SIR}\bigskip
\end{minipage}
\begin{minipage}[c]{1.0\linewidth}
  \centering
  \centerline{\includegraphics[width=.9\linewidth, height=5.6cm]{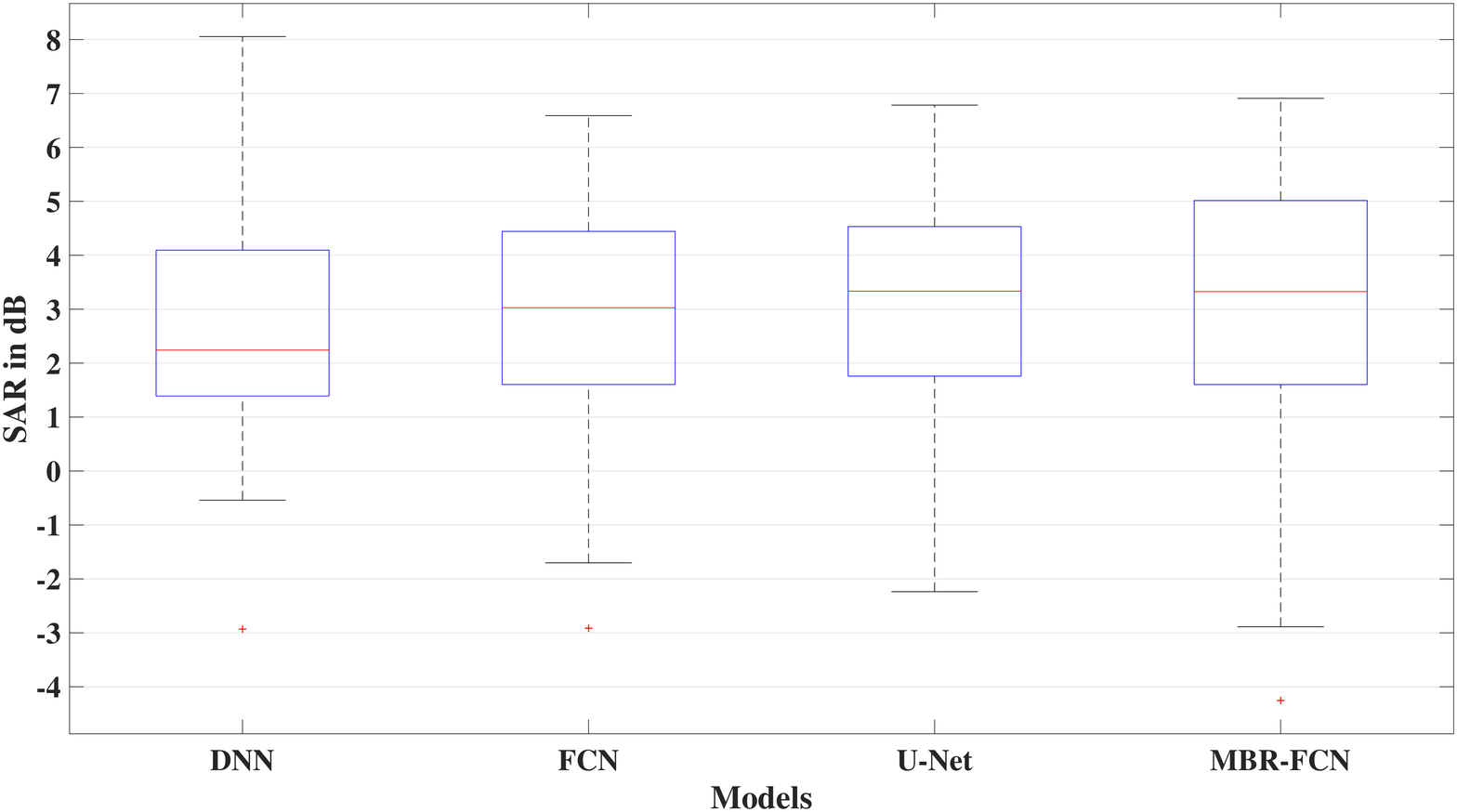}}
  \centerline{(c) SAR}\bigskip
\end{minipage}
\caption{(a) SDR, (b) SIR, and (c) SAR (values in dB) for the separated singing voice of using the following neural networks: deep fully connected (DNN), deep fully convolutional (FCN), the deep U-Net convolutional networks (U-Net), and multi-band multi-resolution fully convolutional neural network (MBR-FCN).}
\label{fig:res}
\end{figure}

The quality of the separated sources was measured using the source to distortion ratio (SDR), source to interference ratio (SIR), and source to artifact ratio (SAR) \cite{vincent:06:pmi}. SIR indicates how well the sources are separated based on the remaining interference between the sources after separation. SAR indicates the artifacts caused by the separation algorithm in the estimated separated sources. SDR measures the overall distortion (interference and artifacts) of the separated sources. The SDR values are usually considered as the overall performance evaluation for any source separation approach \cite{vincent:06:pmi}. Achieving high SDR, SIR, and SAR indicates good separation performance.

Fig. \ref{fig:res}, shows the evaluation results of using the four different neural networks, DNN, FCN, U-Net, and MBR-FCN for SCSVS. For SDR and SIR, the proposed MBR-FCN with very few parameters significantly outperforms all the other models. For SAR, MBR-FCN significantly outperforms the DNN and FCN models. 

The results shown in Fig. \ref{fig:res} were analysed using non-parametric statistical methods \cite{Simpson:16:eassmhdnpsm} to determine the significant differences between the results of the different models. The results of a pair of models are significantly different statistically if $P < 0.05$, Wilcoxon signed-rank test \cite{Wilcoxon:45:icrm} and Bonferroni corrected \cite{hochberg:87:mcp}. For all the measurements (SDR, SIR, and SAR) the results of all the models are significantly different except the following cases: for SDR, there is no evidence of significant differences between the results of FCN and U-Net; for SAR, there is no significant differences between the results of DNN and FCN models, and the results of U-Net and MBR-FCN models.     
\section{CONCLUSIONS}
\label{sec:CONCLUSIONS}
In this paper we introduced a multi-band multi-resolution deep fully convolutional neural network (MBR-FCN) for single channel singing voice separation. The proposed model processes the different frequency bands in the audio spectrogram with different number of filters, different time-frequency resolution, and different scale of dimensionality reduction. The frequency bands with more information are processed with more filters and smaller dimensionality reduction scale than the other bands. Following the constant-Q transform, the MBR-FCN processes the low frequency bands with high frequency resolution filters and the high frequency bands with high time resolution filters. The proposed model with very few parameters significantly out performed the performance of the deep fully connected neural network, the fully convolutional neural network, and the deep U-Net convolutional networks. In our future work, we will further investigate using the same concept of multi-bands multi-resolution to separate various types of music signals from their mixtures. Moreover, we will also combine the concept of multi-band multi-resolution with the concept of gated residual convolution networks \cite{Tan:18:grndcmse} to build more powerfull model for audio source separation.
\ninept
\section*{Acknowledgment}
This work is supported by Cardiff Metropolitan University Research Innovation Award and grant EP/N014111/1 from the UK Engineering and Physical Sciences Research Council (EPSRC).
\vfill\pagebreak


\bibliographystyle{IEEEbib}
\bibliography{strings,refs}
\label{sec:refs}

\end{document}